\documentclass{emulateapj}
\usepackage{lscape}
\begin{document}

\title{ABSOLUTE FLUX CALIBRATION OF THE IRAC INSTRUMENT ON THE \emph{SPITZER
SPACE TELESCOPE} USING \emph{HUBBLE SPACE TELESCOPE} FLUX STANDARDS}

\author{R.~C.\ Bohlin\altaffilmark{1}, K.~D.\ Gordon\altaffilmark{1}, 
G.~H.\ Rieke\altaffilmark{2},
D. Ardila\altaffilmark{4}, S. Carey\altaffilmark{3},
S. Deustua\altaffilmark{1}, C. Engelbracht\altaffilmark{2}, 
H. C. Ferguson\altaffilmark{1}, K. Flanagan\altaffilmark{1}, 
J. Kalirai\altaffilmark{1}, M. Meixner\altaffilmark{1}, 
A. Noriega-Crespo\altaffilmark{3}, K. Y. L.~Su\altaffilmark{2}, and
P.--E. Tremblay\altaffilmark{5}}

\altaffiltext{1}{Space Telescope Science Institute, 3700 San Martin Drive,
Baltimore,  MD 21218, USA}
\altaffiltext{2}{Steward Observatory, University of Arizona, 933 North Cherry
Avenue, Tucson, AZ 85721, USA}
\altaffiltext{3}{Caltech, Spitzer Science Center, MS 220-6, Pasadena, 
CA 91125 USA}
\altaffiltext{4}{Caltech, NASA Herschel Science Center, IPAC/MS 100-22, 
Pasadena, CA 91125 USA}
\altaffiltext{5}{Universit\'e de Montr\'eal, C.P.6128, Succ. Centre--Ville,
Montr\'eal, Qu\'ebec, H3C 3J7, Canada}


\begin{abstract}  
The absolute flux calibration of the \emph{James Webb Space
Telescope} will be based on a set of stars observed by the \emph{Hubble} and
\emph{Spitzer Space Telescopes}. In order to cross-calibrate the two facilities,
several A, G, and white dwarf (WD) stars are observed with both \emph{Spitzer}
and \emph{Hubble} and are the prototypes for a set of \emph{JWST} calibration
standards. The flux calibration constants for the four \emph{Spitzer} IRAC bands
1--4 are derived from these stars and are 2.3, 1.9, 2.0, and 0.5\% lower than
the official cold-mission IRAC calibration of Reach et al. (2005), i.e. in
agreement within their estimated errors of $\sim$2\%. The causes of
these differences lie primarily in the IRAC data reduction and secondarily in the
SEDs of our standard stars. The independent IRAC 8~\micron\ band-4 
fluxes of Rieke et al. (2008) are about 1.5 $\pm$2\% higher than those
of Reach et~al. and are also in agreement with our 8~\micron\ result.
\end{abstract}

\keywords{stars: atmospheres --- stars: fundamental parameters
--- techniques: spectroscopic}

\section{Introduction}

Flux calibrations in physical units for astronomical instruments are required to
make comparisons to physical models of observed objects (Kent et al. 2009). In
particular, one of the main incentives for accurate absolute flux standards is
the requirement for measuring the relative fluxes of redshifted SN Ia spectra in
the rest frame in order to constrain the parameters of the dark energy. These
constraints depend only on the precision of the ratio of fluxes from one
wavelength to another and not on the absolute flux level. Quantitative
descriptions of dark energy are significantly improved when the relative flux
with wavelength is known to an accuracy of 1\% or better (Aldering et al.
2004).

Absolute flux calibrations of spectrometers and photometers are normally derived
from observations of standard stars with well-known spectral energy
distributions (SEDs). For all \emph{Hubble Space Telescope} instruments, all
flux calibrations are traceable to three primary WD standards, G191B2B, GD71,
and GD153. The slopes of these WD SEDs are determined by non-local thermodynamic
equilibrium (NLTE) model calculations using the Hubeny Tlusty Version-203 code
for pure hydrogen atmospheres (Bohlin 2000, Bohlin et al. 2001, Hubeny \& Lanz
1995, Tremblay \& Bergeron 2009). The effective temperature and gravity are
determined by fitting the models to ground-based observations of the Balmer line
profiles (Finley et~al. 1997). 

The absolute flux of the models for these three primary standards is set by
Space Telescope Imaging Spectrograph (STIS) relative spectrophotometry (Bohlin
\& Gilliland 2004, Bohlin 2007a) and the Megessier (1995) absolute flux for
Vega of $3.46\times10^{-9}$~erg cm$^{-2}$ s$^{-1}$~\AA$^{-1}$ at
5556~\AA~(3560~Jy or 3562~Jy for vacuum wavelengths). As discussed in the review
by Hayes (1985), by Megessier (1995), and in \S 4.1.1 below, there is a small
uncertainty in the 5556~\AA\ flux; but this uncertainty just affects the overall
level and $not$ the slope (i.e. ``color'') of the WD models used for \emph{HST}
flux calibrations. Despite suggestions of variations, Hayes (1985) discusses the
evidence for variability of Vega and concludes that the star is likely not
variable. However, Engelke et al. (2010) present evidence for a 0.08 mag
variation of Vega at visible wavelengths.

To compare with previously published calibrations of the \emph{Spitzer Space
Telescope} in the four IRAC bands (Fazio et~al. 2004, Reach et~al. 2005,
hereafter Re05), a set of new observations of white dwarf (WD), A stars, and
solar-analog G stars were made near the end of the cold mission. \emph{Spitzer} 
data in the IRS blue peakup channel (Houck et~al. 2004) or the MIPS 24~$\mu$m
band (Rieke et~al. 2004, Engelbracht et~al. 2007) were included to ensure that
debris disks or red companions do not contaminate the results. These new
observations are supplemented by more data sets for the same stars from the
\emph{Spitzer} archive. Table 1 lists the 14 stars with \emph{HST} based SEDs
that are used for the comparison with the absolute flux calibrations of
Re05 for IRAC, the IRS Instrument
Handbook\footnote{http://ssc.spitzer.caltech.edu/irs/irsinstrumenthandbook/IRS\_Instrument\_Handbook.pdf},
and Engelbracht et~al. (2007) for the MIPS 24~$\mu$m channel. The \emph{HST} 
flux distributions are all in the
CALSPEC\footnote{http://www.stsci.edu/hst/observatory/cdbs/calspec.html/}
database. The Re05 IRAC calibration is based on four A-star SEDs from an
extension of the original Cohen CWW network (Cohen, et~al. 1992a, Cohen et~al.
1999, Cohen et~al. 2003, Cohen 2007), as validated by Price et~al. (2004). Two
of these four primary A star calibrators, HD165459 and 1812095, are included in
Table 1, while our other five A stars are listed as IRAC candidate primary
calibrators by Re05. For a comparison of the \emph{HST} SEDs with the Cohen flux
distributions, see Bohlin \& Cohen (2008), whose minor revisions include average
fluxes that are $\sim$0.5\% lower in the IRAC wavelength range for the set of
seven A stars in Table 1. The K~star calibrators of Re05 are not utilized,
because of the extra complexity of modeling the molecular absorption and because
Re05 used only A~stars to define their final IRAC calibration constants.

In this paper, \S 2 covers the fundamental equations and concept of photometric
flux calibrations, while \S 3 compares the published calibration constants for
four IRAC imaging modes to those that are derived from the \emph{HST} based
SEDs. \S 4 compares our results with the IRAC calibrations of Re05 and with the
8~$\mu$m fluxes of Rieke et~al. (2008, hereafter Ri08). \S 5 includes
suggestions for future efforts to improve the flux calibration accuracy. Finally
in Appendix A, the \emph{HST} method of photometric flux calibration is
illustrated for six \emph{Spitzer} modes. The Appendix is not absolutely
essential to the main thrust of this paper but does expand on several points as
forward referenced in the main body. In addition, the Appendix attempts to
provide a cohesive mathematical foundation for the student or practicioner
of the art of flux calibration.

\section{Generic Calibration Constants}

\subsection{Equations}

This section compares the \emph{Spitzer} flux calibration derived from the
\emph{HST} flux standard stars using the Re05 methodology and nomenclature. For
comparison, Appendix A presents the traditional \emph{HST} flux calibration
methodology. According to Re05, the \emph{Spitzer} flux calibration is defined
such that a point source flux estimate  \begin{equation}\langle
F_{\nu}\rangle=C'N_e=F_{\nu_o}K~,\label{fnu}\end{equation}  where $\langle
F_{\nu}\rangle$ is the mean flux over the bandpass and $C'$ is the calibration
constant for a point source. $N_e$ is the number of detected photo-electrons per
second, either $N_e(pred)$ predicted from the stellar flux and the system
fractional throughput $R$ or $N_e(obs)$ observed in an infinite-radius
photometric aperture. Re05 uses a 10 pixel (12\arcsec) reference radius for the
published calibration constants; but these calibrations refer to surface
brightness (see \S A.2 and Equation (\ref{pnucp})). $F_{\nu_o}$ is the flux at
the nominal wavelength $\lambda_o=c/\nu_o$ for a $\nu F_\nu=constant$ flux
spectrum. $N_e(pred)$ is
\begin{equation}N_e(pred)=A\int{F_\nu\over{h\nu}}~R~d\nu={A\over hc}
\int{F_\lambda~\lambda~R~d\lambda}~, \label{ne}\end{equation} where 
A=4869~cm$^{-2}$ is the collecting area of the \emph{Spitzer} 85cm primary
mirror with its 14.2\% obscuration (Werner 2004).

$K$ is the color  correction 
\begin{equation}K={\int (F_\nu/F_{\nu_o})(\nu/\nu_o)^{-1}~R~d\nu\over
{\int(\nu/\nu_o)^{-2}~R~d\nu}}~,\label{k}\end{equation} where $F_\nu$ is the
actual stellar spectral flux distribution. The nominal wavelength is
\begin{equation}\lambda_o={\int\lambda\nu^{-1}~R~d\nu\over{\int\nu^{-1}~R~d\nu}}
={\int R~d\lambda\over{\int\lambda^{-1}~R~d\lambda}}~,
\label{lamo}\end{equation} where the term after the first equal sign is from
Re05 and the term after the second equal sign is the equivalent formulation of
Hora et~al. (2008).

If the SEDs of the \emph{HST} stars in Table~1 are used for the flux $F_\nu$ to
produce a new calibration constant $C'_{ST}$ and corresponding mean flux 
${\langle{F_{\nu}}^{ST}\rangle}$,  then the ratio of the new to the original
calibration is \begin{equation}{\langle {F_{\nu}^{ST}}\rangle\over \langle
F_{\nu}\rangle}= {C'_{ST}\over C'}= {\int F_\nu/\nu~R~d\nu \over \nu_o~\langle
F_{\nu}\rangle  \int \nu^{-2}~R~d\nu}\label{rat}\end{equation}  or,
equivalently, in terms of integrals over wavelength
\begin{equation}{C'_{ST}\over C'}= {\lambda_o \int F_\lambda~\lambda ~R~d\lambda
\over c~\langle F_{\nu}\rangle \int R~d\lambda}~, \label{ratl}\end{equation}
where $\langle F_{\nu}\rangle$ is the stellar flux derived with aperture
photometry from the \emph{Spitzer} images, as calibrated with the official
calibration constants that appear in the data-file headers. Re05
quotes an uncertainty of 2\% in the IRAC absolute flux calibration. Tests of the
numerical integrations over $\nu$ per Equation (\ref{rat}) or over $\lambda$ per
Equation (\ref{ratl}) are the same to a few parts in $10^5$. 

\subsection{Simplified Concept of a Point Source Calibration}

A specific-intensity calibration for the surface brightness (see \S A.2) of
diffuse sources requires a measure of the total response to a point source in an
infinite aperture, as specified above for $N_e$. However, a flux calibration
that is strictly for point sources has no such requirement and can be explained
simply and elegantly in the case of a stable instrumental configuration with a
linear  response. Stability means that repeated observations produce the same
response, measured in terms of say a background-subtracted net count rate N,
while linearity implies that the count rate is directly proportional to the
physical flux F, i.e. the ratio of flux to count rate will be the same ratio of
F/N over the dynamic range of the system. There is no restriction on the
entrance slit or extraction aperture as long as the same choice is made for both
stars and the extracted count rate is repeatable for both stars. The measured
count rate can be in a certain radius aperture for point source photometry or of
a certain height for a resolution element of a spectrophotometer. If one star is
a flux standard with known flux, then the ratio F/N defines a point source
calibration constant P, so that the second star with unknown flux has the same
constant measured ratio; and the unknown flux is simply F=PN. This constant P
might be alluded to as a sensitivity but is really more properly an inverse
sensitivity, because a more sensitive instrument will have a higher count rate
for a source of the same flux.

The main complication of this concept is due to the different spectral
resolutions of the flux standard and the unknown star. A common example is a
standard star with a tabulated medium resolution SED. For broadband photometry,
the average flux of the standard over the bandpass must be calculated as in
Equation (\ref{favl}) or (\ref{favnu}), which is straightforward if the spectral
resolution is much better than the band width. In the case of a spectrometer
calibration with a resolution that is lower than the tabulated resolution of the
standard, the calibration P is defined simply as the known SED, F, binned to the
bandpass of the instrument to be calibrated divided by the response spectrum N
for the same standard star. More properly, P as a function of wavelength is
defined as the convolution of the known SED, F, with the instrumental
line-spread function, LSF, divided by the count rate spectrum, N, of the
standard convolved with the LSF of the standard star spectrum, which brings the
numerator and denominator spectra of P to the same resolution and enables a
pixel-by-pixel division of F by N after resampling to the same wavelength scale.
This procedure may fail for the case where a low resolution standard star SED
must be bootstrapped to a calibration of a much higher resolution spectrometer
where the sensitivity of the high resolution data changes significantly over the
resolution element of the known SED. For example, a single echelle order may
have a variation in sensitivity by a factor of 10 or more over a wavelength
range covered by only one or a few resolution elements of the flux standard.

\section{Spitzer Calibration}

\emph{Spitzer} observations of the sources were taken either as part of a cycle
5 Director's Discretionary Time program (PI: Gordon) or from existing archival
observations.  The stars were observed in the four IRAC bands, and as
many as possible were observed in the IRS blue peakup band or MIPS 24~\micron\
band. The main goal of the IRS blue peakup and MIPS 24~\micron\ observations is
to check for dust emission (e.g., a debris disk) or faint red companions.  

\subsection{Data Reduction}

All the IRAC and IRS blue peakup data (reduction version S18.7.0) were
downloaded from the \emph{Spitzer} archive. The photometry is measured using a 3
pixel radius aperture and sky annulus with radii of 10 and 20 pixels on each
individual image. As our sources are faint, refined positions are determined by
centroiding on the star in each observation mosaic image. An aperture of three
pixels radius is chosen in order to minimize noise from the sky and
contamination from other sources in our relatively crowded fields. To get the
total stellar flux, $\langle F_{\nu}\rangle$, our 3 pixel radius aperture
photometry is corrected to the standard calibration aperture sizes per Table~2. 

Table~2 includes the nominal wavelengths computed from Equation (\ref{lamo}),
the size of the standard reference aperture for each filter (i.e. 10 pixels for
IRAC and infinite for IRSB and MIPS), the aperture correction needed to convert
our three pixel photometry to the reference aperture size, and the references
for the aperture corrections and instrumental-throughput spectral-response R vs.
vacuum wavelengths. Our aperture correction values for the IRAC photometry of
HD165459 agree with the tabulated values from Hora et~al. (2008) to better than
0.25\%, even though the Hora background annulus is 10--20 pixels instead of the
12--20 pixels used by Re05 for the standard 10 pixel photometry. The IRAC
aperture correction from 10 pixels to infinity is discussed in \S A.2. Our
nominal vacuum wavelengths $\lambda_o$ are within 0.4\% of those in Re05 and
within 0.2\% of Hora et~al. (2008).

\begin{figure}
\epsscale{1.2}
\plotone{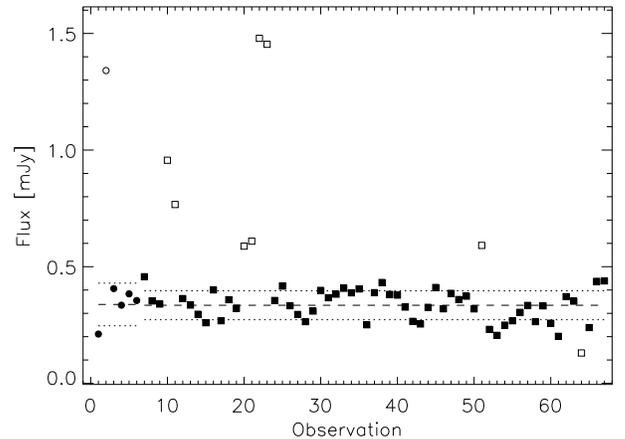}
\caption{The IRAC4 photometry for G191B2B for the 1st (circles) and
2nd (squares) AOR.  The filled symbols gives those
measurements that were used in computing the averages (dashed lines)
and standard deviations (dotted lines).  The open symbols are the
measurements that were iteratively sigma-clipped because of contamination
by cosmic ray hits.
\label{fig1}}  
\end{figure}

For G191B2B, a nearby bright star produces an artifact in its sky region; and
the pixels affected are rejected prior to determining the sky flux.  For each
independent observation (Astronomical Observation Request, AOR), the photometry
from the multiple image frames is averaged after sigma-clipping rejection of
outlying points. For example, Figure~\ref{fig1} illustrates
the rejected and accepted IRAC4 photometry points for the two AORs for G191B2B. 

The weighted average IRAC and IRS fluxes for each star and band are given in
Table~3 after multiplication of our three pixel radius photometry by the
aperture corrections in Table~2. In order to achieve robust results, IRAC
observations with a signal-to-noise ratio (S/N) of less than 7 or with a
location more than 25 pixels from the center are rejected and do not contribute
to the averages in Table~3. The restriction to centrally located sources avoids
any confusion due to possible errors in the flat fielding procedure. The second
line for each star in Table~3 is the synthetic photometry predictions computed
from our standard star SEDs per Equation (\ref{favnu}).

\subsubsection{Uncertainties}

\begin{figure}
\epsscale{1.2}
\plotone{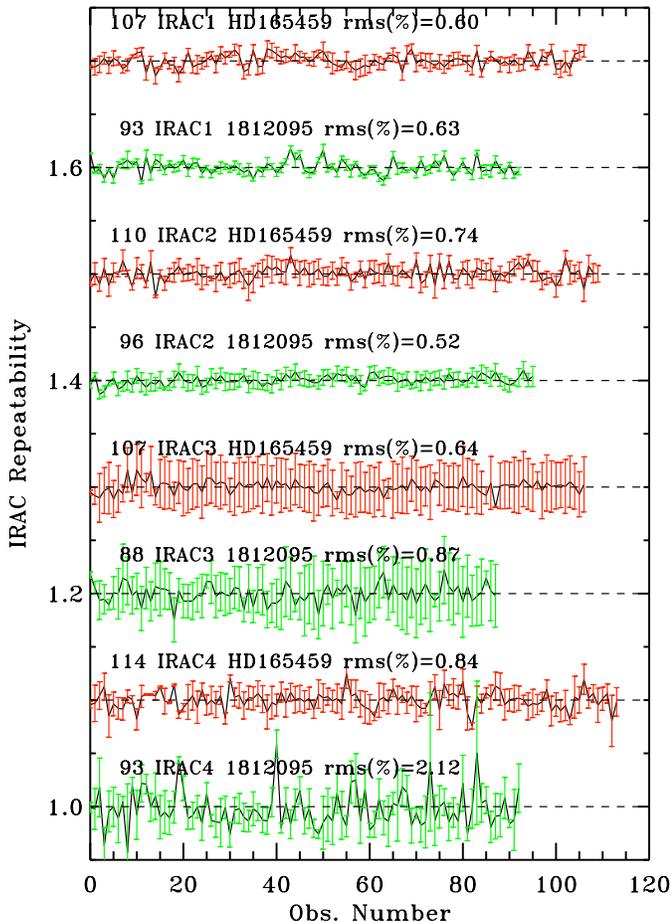}
\caption{
Observations of two Re05 primary stars in the four IRAC bands, where
each point is the result from one AOR. Each set of points is offset by 0.1 along
the Y-axis from the set below. The points displayed all have a S/N of at
least 7 and are within 25 pixels of image center in each IRAC channel. The
red points are for HD165459, while 1812095 is shown in green. Each of the eight
panels is labeled with the number of observations (AORs), the IRAC band, the
star name, and the rms of the points shown.
\label{fig2}}
\end{figure}

\begin{figure}
\epsscale{1.15}
\plotone{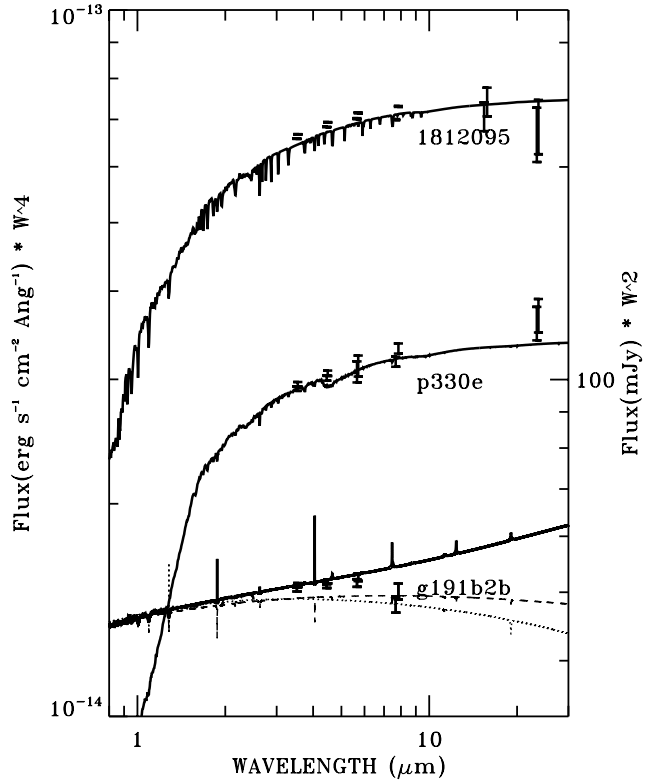}
\caption{
The continuous lines are SEDs typical of our three spectral categories of
standard stars scaled by $\lambda^4$ on the left axis and by $\lambda^2$ on the
right axis with $\lambda$ in \micron. The absolute fluxes are multiplied by a
factor of two for G191B2B and 1812095. The statistical error bars of
$\pm1\sigma$ are shown for the measured Spitzer photometry at the nominal
wavelengths and at the effective wavelengths. For G191B2B, a Tlusty LTE model
(dash) and a NLTE 60,000K model (dots) with solar CNO (Gianninas 2010) are shown
in addition to the standard NLTE model (solid). \label{fig3}} 
\end{figure}

Many ($\sim$100) AORs exist for the four primary Re05 standard stars. For
example, Figure~\ref{fig2} shows the observations of two of our stars,
HD165459 and 1812095, that are also Re05 primary standards. Each point in
Figure~\ref{fig2} represents the sigma-clipped average of multiple image
frames in one AOR. The rms indicated on each panel is the scatter among the
remaining AOR observations after rejecting points that deviate by more than
3$\sigma$ from the average. The brighter star HD165459 has 2, 5, 4, and 1
rejected deviant AORs with an rms for the remaining points of 0.6, 0.7, 0.6, and
0.8\% for channels 1--4, respectively, while the comparable rms dispersions from
Re05 are 1.7, 0.9, 0.9, and 0.5\%. In order to compute the weight for each AOR
included in the average fluxes of Table 3, our repeatabilities for HD165459 are
added in quadrature with the statistical uncertainty for each independent
observation. Figure~\ref{fig3} compares the Spitzer broadband fluxes to
our absolute flux distributions for one example of each of our three stellar
classes. The differences between the flux levels for the nominal and effective
wavelengths quantify the ambiguity associated with assigning monochromatic
wavelengths to the broadband \emph{Spitzer} photometry. 

\subsubsection{Comparison with the Re05 Photometry}

Re05 based the final, recommended IRAC calibration on four primary A-star
standards, two of which, HD165459 and 1812095, are among our standard stars. 
Table 1 of Re05 contains the photometric fluxes for these two stars along with
their other two primaries, HD180609 and BD+60$^{\circ}$1753. On average, our 
extracted IRAC photometric fluxes are 2.5, 2.8, 2.4, and 1.1\% higher than
the corresponding Re05 tabulations for channels 1--4, respectively.

Our IRAC photometry from each image is corrected for distortions and pixel phase
as recommended by Hora et al. (2008). These corrections differ from Re05, who
used a preliminary version of the Hora et al. work. Because of these different
data reduction procedures, our corrected photometry is expected to be
systematically brighter than Re05 by 1.1, 1.4, 0.6, and 0.5\%, for bands 1--4,
respectively.  These expected systematic difference between Re05 and our
photometry account for around half of the actual differences. The remaining
unexplained differences of up to 1.8\% for IRAC3 must be due to the changing
IRAC pipeline processing and/or a somewhat different selection of IRAC
observations to include in the average for each star.

\subsubsection{Special Cases}

There are no IRAC4 measurements for 1732526. Two additional \emph{HST} standard
G stars have IRAC observations (C26202 and SNAP-2), yet neither star has high
enough quality observations to be included in  this work.  C26202 is in the
CDF-S (Smith et~al. 2003) and has a cooler companion at 3--4\arcsec\ that
produces blended IRAC images and precludes accurate photometry of the separate
stars.  The second G star, SNAP-2, has IRAC data but lies off-center by $\sim$50
pixels where flat fielding errors might be important.

\subsubsection{MIPS}

The raw MIPS data were downloaded from the Spitzer archive and reduced using the
MIPS Data Reduction Tool (Gordon et al. 2005). In addition, several additional
steps to remove residual instrumental signatures are used (see Engelbracht et
al. 2007 for details). Given the crowded nature of some of the fields, PSF
fitting code is required to extract the MIPS24 photometry of our sources; and
our choice is StarFinder (Diolaiti et al. 2000).

\subsubsection{Predicted Fluxes}

The measured and synthetic fluxes appear in alternate rows in Table~3. For the
three pure hydrogen WDs, the temperature and gravity of the TLusty NLTE models
are derived from fits to ground-based spectra of the Balmer lines. For the pure
He WD LDS749B, the model of Bohlin \& Koester (2008) is used. For the A stars,
Bohlin \& Cohen (2008) fit NICMOS spectrophotometry from 0.8--2.5~\micron\ and
ground-based photometry with Castelli \& Kurucz (2004, hereafter CK04) model
SEDs. Similarly for the G stars, Bohlin (2010, hereafter B10) fit STIS and
NICMOS spectrophotometry. Comparing our Table 3 synthetic fluxes with the
corresponding values in Table~6 of Re05 for the two stars in common confirms
that Re05 used stellar SEDs without modification from the Cohen CWW network. A
generous statistical uncertainty of 1\% is assigned to our Table 3 synthetic
fluxes to account for the effects of the broad band Poisson noise and
repeatability of STIS spectrophotometry (Bohlin 2002) on the fitting of models
to the \emph{HST} spectrophotometry. More details of the \emph{HST} synthetic
fluxes in Table 3 and their systematic uncertainties are discussed in \S 4.1.

\subsection{Linearity}

\begin{figure} 
\epsscale{1.1}
\plotone{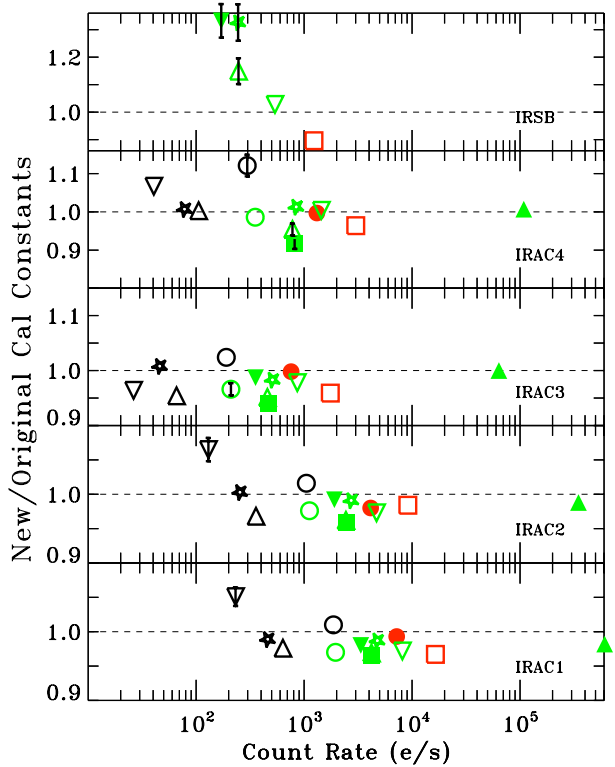}
\caption{
Check of the linearity of the observed IRAC and IRSB photometry versus source
brightness. The
plotted ratios per the equivalent Equations (\ref{rat}) and (\ref{ratl}) are a
measure of the mean flux of the \emph{HST} SEDs divided by the measured
photometric flux per the \emph{Spitzer} pipeline calibrations. The stars are
color coded by spectral type, with black, green, and red for WDs, A~stars, and
G~stars, respectively. One sigma error bars are shown on the points that differ
from unity by more than 3$\sigma$. The key for each symbol type is at the right
side of Figure~\ref{fig5}. Notice the progressively more
compressed scales in the upper three panels. \label{fig4}} 
\end{figure}

The linearity of five of the \emph{Spitzer} detector systems can be evaluated by
comparing the observed photometry to the actual stellar brightness. A more
common linearity check is a comparison of count rate vs. well depth. For the
sixth system, MIPS24, our three data points are insufficient to reach any
conclusion; see Engelbracht et~al. (2007) for details of the confirmation of
MIPS 24\micron\ linearity. Using the modeled \emph{HST} SEDs as the stellar
flux, Equation (\ref{rat}) is the ratio of predicted to measured fluxes; and
that ratio is shown in Figure~\ref{fig4} as a function of stellar
brightness for our three classes of stars. IRAC3 has the narrowest bandpass and
the lowest countrates. G191B2B is significantly high in the IRAC4 band; but that
one anomalous point is not an indication of non-linearity. However for IRSB,
even over the small dynamic range of 7, the five data points show evidence of
some issue with the photometry. The ratio for the faintest IRSB source 1732526
is 30\% high, while the brightest source P041C is 10\% low. However, Gilliland
\& Rajan (2011) discovered that P041C is a double star with an M star companion
separated by 0.57\arcsec, which could make the observed IRSB flux too high by
$\sim$10\%. These problems preclude an accurate comparison of the \em{HST}
versus \em{Spitzer} IRSB calibration.

\subsection{Dust Rings}

Many stars are encircled by a ring of cool dust that emits strongly at longer
wavelengths. These sources can be identified by small ratios of predicted to
observed flux that fall off-scale on Figure~\ref{fig4}. Our observed
excess flux at 24~\micron\ for HD165459 confirms the evidence for a disk found
by Rieke et~al. (2005), who reported a 40\% excess.
Su et~al. (2006) revised the excess to 46.7\%.

\begin{figure} 
\epsscale{1.1}
\plotone{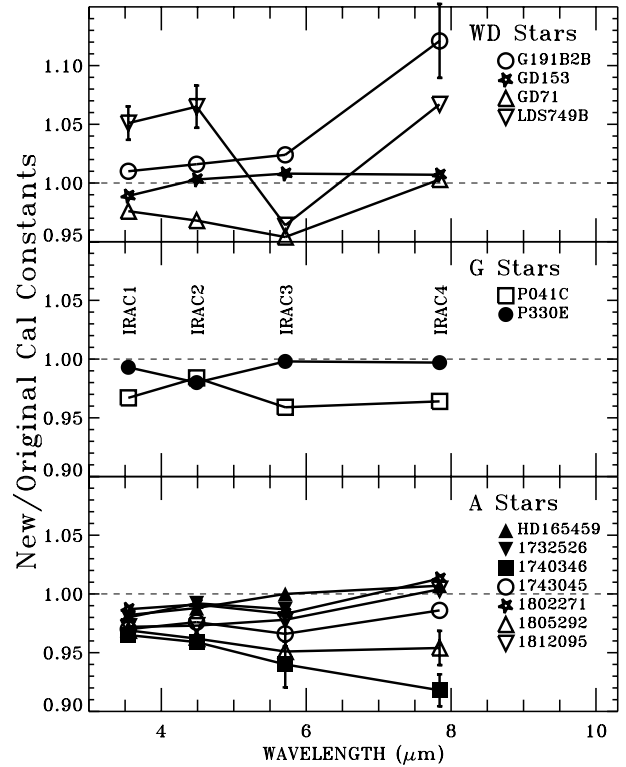}
\caption{
Ratios as in Figure~\ref{fig4} of calibration constants computed from
the \emph{HST} based fluxes to the IRAC values of Re05. WDs are in the top
panel, solar analogs are in the center panel, and A~stars are at the bottom. One
sigma error bars are shown on the points that differ from unity by more than
3$\sigma$. The key to the individual symbols is at the right side of each
panel.The scales are the same for all three stellar types, but there is an
offset of 0.05 in the top panel. \label{fig5}}  
\end{figure}

Our excess of a factor of 2.5 for 1740346 in the 16~\micron\ band suggests the
presence of a contaminating debris disk. While there is no evidence of dust
emission below 8~\micron\ for HD165459, the low values for 1740346 in
Figure~\ref{fig5} suggest the presence of
hotter dust than is around HD165459. Therefore,
1740346 is not included below in the comparison with the Re05 calibration of
IRAC.

\subsection{Results}

Per Equation (\ref{rat}) or (\ref{ratl}), Figure~\ref{fig5}
compares the \emph{HST} calibration constants
to those published by Re05. One $\sigma$ error bars appear on the points that
differ from unity by more than the 3$\sigma$ uncertainty of the ratio. The 
problematic IRSB data are not shown; and the MIPS24 ratios for the three stars
without dust rings are insufficient to draw any firm conclusions, although all
three MIPS ratios are within $\sim$10\% of unity.

With a 4$\sigma$ statistical significance, the IRAC4 ratio of 1.12
for the observation of G191B2B is the largest deviation from unity in
Figure~\ref{fig5}. As a check on this problematic case, the
background level in the IRAC4 band was studied by constructing a histogram of
the set of all three-pixel radius photometry in a 120 pixel square, centered on
G191B2B. There are no background regions in the vicinity of the star that are
low enough to bring the observed result to the predicted level with a ratio of
unity in Figure~\ref{fig4}; and there is only $\sim$1\% probability of
a sky level that would reduce the discrepancy to 2$\sigma$.

Table 4 reports the average results separately for the WDs, for two Re05 primary
A stars, and for the group of A and G stars.  These values are the weighted
averages of the ratios of the synthetic to the measure fluxes in Table 3, where
the uncertainty for each star is the measurement error in Table~3 combined with
the 1\% uncertainty assigned to the synthetic values. These uncertainties are
the statistical errors and do not include any possible systematic errors that
would affect all stars equally. As a group, the averages for the WDs are
significantly above the other averages, especially with the inclusion of the
IRAC4 ratio for the hottest WD, G191B2B, that is 12\% high in
Figure~\ref{fig5}. Individually, many of the WD results could be
explained as not significant at the $3\sigma$ level or because of the less well
vetted pure helium model for LDS749B (Bohlin \& Koester 2008).

NLTE effects in the IR become more pronounced at higher stellar temperatures
(Bohlin 2000), eg. the pure hydrogen LTE model for G191B2B shown in
Figure~\ref{fig3} falls near the IRAC4 data. For G191B2B, metal
line-blanketing is observed at the 1--2\% level in the FUV (Barstow et~al.
1999). Per Gianninas et~al. (2010), an additional NLTE model spectrum at 60,000
K and $\log g= 7.5$ with CNO metals at the Asplund et~al. (2005) solar abundance
also appears in Figure~\ref{fig3} near the LTE model. This metal
abundance is not the actual metallicity but is only a proxy for the effect of
the metals, which is to reduce the NLTE effects and cool the upper atmosphere,
where the IR continuum is formed. There is no coherent simultaneous
determination of the metal abundances along with $T_\mathrm{eff}$ and $\log g$
from the Balmer lines (Barstow et~al. 2003); however, the abundances suggested
by Barstow et~al. (2003) are considerably less than the solar Asplund et~al.
(2005) values, so that a model with proper trace metallicities for G191B2B in
the IR should fall somewhere between the 60,000K CNO model and the pure hydrogen
NLTE model, potentially within 1--2$\sigma$ of all the IRAC fluxes. 

The significant deviation from unity in
Figures~\ref{fig4}--\ref{fig5} for G191B2B should be
explored with a proper model. Meanwhile, the most relevant result in Table 4 is
for the set of eight G and A stars, for which our calibration constants are
lower than Re05 for IRAC1--4 by 2.3, 1.9, 2.0, and 0.5\%, respectively. For the
four Re05 primary stars, our re-reduced photometry is 2.5, 2.8, 2.4, and 1.1\%
higher than Re05. Higher extracted photometry implies lower calibration
constants; and the measured photometry differences correspond to the calibration
constant differences to an accuracy of better than 1\%. Understanding that our
differences with Re05 are mostly due to differences in the photometry extracted
from the IRAC images helps verify that both calibrations have been properly
derived per the adopted common methodology and that our results are expected to
be lower by about the amounts computed in Table 4 for A+G stars.

While the calibration constants themselves depend directly on the accuracy of
the bandpass throughput $R$ per Equation (\ref{favl}) or (\ref{favnu}), our
methodology and that of Re05 involve the same integral of the stellar SED over
$R$, so that errors in $R$ cancel to first order in the comparison of the two
sets of results. However, bandpass errors, such as an overall shift of $R$ in
wavelength, changes the predicted $N_e(pred)$ per Equation (\ref{ne}). For
example, a shift of +0.1~\micron\ in the 2.9~\micron\ wide IRAC4 band, i.e. a
shift in wavelength by 3\% of the band width, would cause a 3.8\% decrease in
$N_e(pred)$ for the WD, A, and G stellar types. Per Equation (\ref{ne}), such a
3.8\% error would be difficult to distinguish from a simple 3.8\% compensating
error in the laboratory measurements of $R$, because all of our standard stars
have nearly the same Rayleigh-Jeans slope in the IR. A bandpass shift would only
be important in the case of strong spectral features within the bandpass or in
the case of a source with an SED much cooler than our A and G standard stars.

\section{Details of Absolute Flux Calibration}

\subsection{Details of the \emph{HST} Calibration}

The discussion of our measures of instrumental response (N) appear above in  \S
2.1--2.2, while this section covers the details of the flux (F) in the
calibration constant P=F/N.

The essence of the \emph{HST} flux system is to establish standard star SEDs and
then measure other stars relative to these standard candles. Pure hydrogen WD
stars are chosen as these \emph{HST} fundamental standard candles, because their
atmospheric  models are simpler than other stars where metal lines, molecular
lines, and convection add complications. The models of our pure-hydrogen primary
WD standards are specified by two parameters $T_\mathrm{eff}$ and $\log g$, both
of which are defined by the Balmer line profiles. To establish absolute flux
standards, the unreddened WD stars G191B2B, GD71, and GD153 are observed with
STIS, which also observed Vega in the same modes with the same dispersion.
Because STIS is precisely linear even into the regime of many times
over-saturated in its CCD (Gilliland et~al. 1999, Bohlin \& Gilliland 2004), the
relative flux between Vega and each WD is measured (Bohlin \& Gilliland 2004);
and the well-known absolute monochromatic flux at 5556\AA\ for Vega establishes
the absolute flux of the model SED for each WD. Minor complications arise due to
the slowly changing STIS sensitivity with time and the gradual loss of charge
transfer efficiency in the CCD. These corrections are tracked to better than
1\%, and the STIS response is corrected for these effects as a function of
wavelength.

Observations of the three fundamental primary WDs establish the instrumental
flux calibration. For example, a set of observations in the low dispersion modes
of STIS below 1~\micron\  and from 0.8--2.5~\micron\ in the grism modes of
NICMOS establish the flux calibrations and enables the creation of secondary
flux standards. Bohlin \& Koester (2008) demonstrated that such STIS and NICMOS
spectrophotometry of LDS 749B, a pure helium star, could be modeled to the
statistical precision of the data. NICMOS spectrophotometry, supplemented by
ground-based photometry for the A stars used in this paper was modeled by Bohlin
\& Cohen (2008) with $T_\mathrm{eff}$, $\log g$, $\log z$, and color excess
$E(B-V)$ as free parameters. Similarly, B10 modeled STIS and NICMOS fluxes for
the G stars discussed here. The BVR photometry used to define the A-star models
is from Mount Hopkins Observatory (Cohen et~al. 2003) and is referenced to the
9400K Vega model and to Vega photometric values from Maiz Apell\'aniz (2007).

\subsubsection{Uncertainties}

Establishing proper uncertainties is a complex process, involving the division
into categories of systematic and statistical errors, which can each be divided
into  sub-categories of absolute level and relative slope of flux vs. wavelength
(i.e. ``color''). In many cases, the statistical scatter can be reduced below
the systematic error bars by repeated observations or by binning spectra.
Systematic uncertainty of the \emph{HST} WD flux scale is estimated as $\sim$1\%
by B10 for the ratio of the flux at 5556\AA\ to fluxes in the 1--2.5 \micron\
range. The total systematic uncertainty in our IR flux scale should be less than
2\%, even when the 0.7\% uncertainty in the absolute 5556\AA\ flux (Megessier
1995) is included. Considerable confidence in the estimate of precision in the
slope of the relative flux with wavelength can be gained by examining the
internal agreement among the \emph{HST} standards:  i) Bohlin (2007a)
illustrates the $<<$1\% agreement of the three primary WD models with their
calibrated fluxes, which means that if the slope of any model, which depends
primarily on $T_\mathrm{eff}$, is in error, then the other two model
temperatures must be similarly in error in order to make the same change in flux
vs. wavelength for all three stars from 0.12--2.5~\micron.  ii) The NICMOS
photometric calibration is based on the \emph{HST} primary standard G191B2B and
the secondary standard P330E per de Jong (2006), where the consistency of the
absolute spectrophotometry is confirmed for these two SEDs from the
CALSPEC\footnote{http://www.stsci.edu/hst/observatory/cdbs/calspec.html/}
database. In particular, de Jong's Figure 2 shows an agreement of 0.8--1.4\%
between the NICMOS calibrations derived from the SEDs for P330E and G191B2B over
the combined 0.9--2.4 \micron\ range of the three cameras. Finally, the
extrapolation of the \emph{HST} NICMOS fluxes from 2.2~\micron\ to the first
IRAC band at 3.6~\micron\ is rather independent of the particular stellar model.
For example, among the seven A stars in Table 1,  Bohlin \& Cohen (2008)
demonstrate that all the measured NICMOS SEDs fit their model SEDs over the
0.8--2.4 \micron\ range within 1\% in broad wavelength bins; and the maximum
difference in the 2.2/3.6~\micron\ ratio is 0.7\% between the minimum
$T_\mathrm{eff}$=7650K model for 1743045 and the maximum 9100K model for
1802271.

\subsection{Details of the Re05 Calibration}

As discussed above, our measured IRAC instrumental responses are smaller than
published by Re05 for the same stars. However, Re05 used data from the IRAC
pipeline processing version S10, while version S18.7.0 is utilized for this
paper. To be directly comparable, any two independent calibrations should use
the same pipeline processing version, as well as the same algorithms for
extracting the point source fluxes from the images.

As discussed above and in the Appendix, the equations that define our new
calibrations are exactly equivalent to those of Re05, so that differences arise
only from different measured instrumental photometry or from differences in the
adopted stellar fluxes.

The stellar fluxes used by Re05 are from the CWW grid, while 
the revised SEDs from Bohlin \& Cohen (2008) are used here. Some major
differences in the derivation of model SEDs are that Bohlin \& Cohen fitted
models to NICMOS spectrophotometry in the 0.8-2.4~\micron\ range rather than to
the 2MASS plus I band photometry used for determining the CWW SEDs. The
CK04 model grid was used to fit the data;
and rather than specifying the model $T_\mathrm{eff}$ and $\log g$ from spectral
classification spectra, Bohlin \& Cohen found the best fitting model from the
grid, allowing $T_\mathrm{eff}$, $\log g$, $\log z$, and the reddening E(B-V) to
vary as free parameters. Despite these differences, the resulting SEDs differ
only slightly from the CCW fluxes used by Re05. For example over the IRAC
3--9~\micron\ range, our SEDs for the IRAC primary stars HD165469 and 1812095
agree with the CWW SEDs used by Re05 to 0.5 and 1--2\%, respectively.

\subsection{Details of the Ri08 Calibration}

Ri08 start by determining a best calibrated value for an equivalent Vega
photospheric flux at 10.6 \micron\ and use a Vega theoretical model normalized
to this value to predict the photospheric flux density at 2.22 \micron. This
prediction is robust against different models for A0 star photospheres. After
correction for the small contribution for the extended debris ring found in
interferometry, the resulting prediction is compared with measurements of the
absolute flux density from Vega near 2.22 \micron. The agreement is good; and
the two approaches are averaged to generate a 'best' A-star-based calibration at
2.22 \micron. Ri08 independently generated a calibrated solar spectrum as a
combination of the measurements of Thuillier et al. (2003) out to 2.4 \micron\
and an Engelke function from 2.4 to 12 \micron\ (over which range the Engelke
function gives a good fit to a number of accurate calibrated solar
measurements). Ri08 use this spectrum to compare with the observed K-[8] color
of solar-type stars and extrapolated the result to 10.6 \micron\ for an
independent test of the beginning calibration at this wavelength. Their
calibration makes no reference to the visible calibration, although  the
predicted color of Vega based on the absolute calibrations at V and K is
consistent with the observed color of the star to within about 2\%. The V-K
colors of solar-type stars are also consistent roughly within this error with
their calibration at 2.22 \micron\ using the Thuillier measurements to translate
to V.  The Ri08 results should be understood as a purely infrared-based
calibration with an estimated accuracy in this region of 2\%.

The Ri08 calibration suggests that the IRAC4 8~\micron\ fluxes should be 1.5
$\pm$2\% higher, while our fluxes are  $\sim$0.5\% lower than Re05. Thus, our
results agree with Ri08 within the uncertainties. A new extraction of the IRAC
photometry for the 32 Ri08 stars compared with the Ri08 tabulation of predicted
fluxes produces a mean predicted-to-observed ratio that is 2.1~$\pm$0.5\% above
unity, i.e. confirming the Ri08 difference of 1.5\% within the uncertainties.

\subsection{Comparison of Methodologies}

The following points compare the Ri08 technique to the similar \emph{HST} and
CWW methods for establishing standard star SEDs.

i) Ri08 establish two independent IR absolute flux determinations for A stars
	(Vega) and for G stars (the Sun), for which agreement provides a
	powerful confirmation of the results. Both the \emph{HST} and CWW
	absolute levels are traceable only to the visible range for Vega.

ii)  In the Ri08 method, a large number of flux standards are more easily and
	cheaply established from existing accurate photometric systems, so that
	statistical uncertainties can be reduced well below systematic
	uncertainties. The other methods are more observationally expensive,
	especially in the case of the \emph{HST} flux measurements.

iii) Because of the similarity of SEDs at long wavelengths, errors between
	widely separated IR wavelengths are minimal. Ri08 normalized the adopted
	SEDs to IR absolute flux measurements, so that errors in the IR are
	minimized. 
	
iv) The shape of stellar SEDs short of 1.5~\micron\ becomes increasingly dependent
	on the stellar temperature and on the metallicity.
	Thus, the uncertainty of the flux calibration below 1.5~\micron\ grows
	with decreasing wavelength making the Ri08 technique less useful for
	calibration in the short wavelength (0.6-1.5~\micron) range of JWST.

\section{Recommendations}

Several deficiencies in our prototype set of flux standards must be addressed
before \emph{JWST} can be calibrated to our goal of 1\% precision at
0.6--30~\micron. Required improvements include i) better stellar
atmosphere grids and a better understanding of their uncertainties, ii) updated
and expanded lists of potential standard stars in each of our three spectral
type classes, and iii) resolution of the proper normalization of these models to
measures of absolute stellar fluxes in the visible and IR.

\subsection{Model Atmosphere Grids}

Perhaps, the best way to assess systematic errors within a set of similar type
stars is to compare independent sets of models to the measured flux
distributions, which B10 did for the G stars by fitting both MARCS models
(Gustafsson et al. 2008) and those of CK04. In broad continuum bins,
both sets of models agree with the \emph{HST} fluxes to $\sim$0.5\%. 
However, both of these sets have serious deficiencies. The MARCS grid does not
extend past 20~\micron, while the CK04 grid has rather coarse wavelength spacing
with a total of 1221 points and only one point between 10 and 40~$\mu$m. A third
independent model grid, an update of the CK04 models with good wavelength
resolution, and an extension of the MARCS grid beyond its current 20~\micron\
limit would be ideal. Shortward of 20~\micron, the best-fit CK04 models agree
with the best-fit MARCS models to 1\% for the G stars in broad continuum bands.
Combining this 1\% with a 1\% systematic uncertainty in the \emph{HST} WD flux
scale results in the B10 estimate of a possible systematic 2\% uncertainty of
the broadband IR fluxes of individual G stars with respect to the V band. 

For the A stars, a second independent model grid is needed, because the MARCS
models are limited to maximum temperatures of 8000K. Both the original CWW SEDs
and the newer Bohlin \& Cohen (2008) SEDs are extrapolations of the measured
fluxes into the IR using models that are based on computer code traceable to R.
Kurucz. The best check on these IR fluxes of the A stars was from the two models
provided by T. Lanz at 9400~K and 8020~K. While the agreement of the Lanz SED
with CK04 is within 2\% for the 9400~K model, the Lanz SED is brighter than CK04
by 4\% at 10~\micron\ for the 8020~K model, which raises the question of the
accuracy of model SEDs, in general. Because of the scatter among the WDs and
because only two G stars are included, our results are based
mainly on A stars. Furthermore, the final Re05 IRAC calibration is based
entirely on A stars, so that more work on independent A star grids is
essential. 

\subsection{Sample of Stars}

Our goal of comparing results among different stellar types is compromised by
the large scatter and offset of the average results for the WDs and by the poor
significance of the G-star result that is based on only two stars. Perhaps,
warm-mission IRAC1 and IRAC2 observations of our four WD stars could illuminate
the reality of their large scatter in the top panel of
Figure~\ref{fig5}. Among the A stars,
Figure~\ref{fig5} shows the most scatter at 8~\micron, where
1740346 is excluded because of evidence for dust at 16~\micron. However, 1805292
has a 3$\sigma$ deviation without a strong indication for dust. Thus, more A
stars would reduce uncertainties, especially with respect to the pure  A-star
calibration of Re05. The remaining two primary and primary-candidate A star
standards of Re05 should be observed with STIS and modeled to achieve a more
complete and precise comparison between the \emph{HST} and Re05 A-star results.

In order to confirm the A star comparison with IRAC, more G stars and more WD
stars with good existing cold mission \emph{Spitzer} observations are needed.
Plenty of observations of brighter stars with both IRAC and MIPS 24\micron\ exist
in the Spitzer archives, except for WDs. Thus, the WD category will be
supplemented by late O or B type stars. This new set of standards is also
needed to cover the large dynamic range and spectral variety required to
calibrate the full suite of JWST instrumentation but will require new STIS
spectrophotometry to provide the link to \emph{HST}. ACS photometry, WFC3
photometry, and WFC3 grism spectrophotometry will also help in firmly
establishing the tie to the \emph{HST} flux scale, especially for the WFC3 data
at wavelengths longward of the STIS cutoff at 1\micron.

\subsection{Absolute Flux Level}

\begin{figure}
\epsscale{1.1}
\plotone{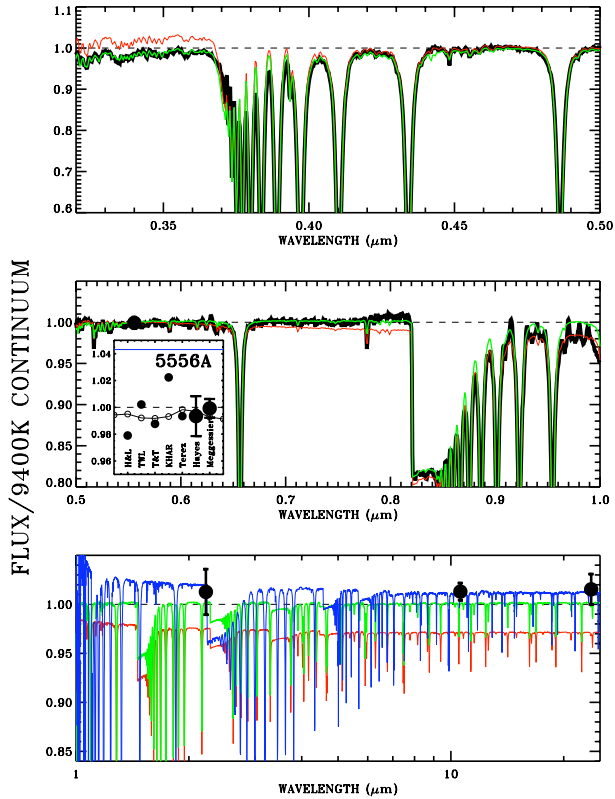}
\caption{
Ratio to the 9400~K model $continuum$ level from  0.32--30~\micron~in three
segments for the STIS measured fluxes (black), for the Kurucz
9400~K model (green), for the Kurucz 9550~K model (red), and for the Ri08 fluxes
(blue). In the top two panels, the STIS spectrophotometry (black line) is mostly
hidden under the green line. The three monochromatic, absolute IR-flux values of
Ri08 appear on the blue Ri08 SED as filled circles with error bars. The
Megessier flux value at 0.5556~$\mu$m is at unity, where the 0.7\% error bar is
inside the circle. The inset is a blowup of this 5556~\AA\ region that shows
this point along with the robust set of direct absolute flux measurements
summarized by Hayes (1985). The five smaller filled circles are labeled per the
Hayes nomenclature and are averaged to get his estimate of
$3.44\times10^{-9}$~erg cm$^{-2}$ s$^{-1}$~\AA$^{-1}$ at 5556~\AA\ with its
1.5\% error bar that is labeled as Hayes in the inset graph. Also in the inset,
the nine small open circles connected with a line are the Hayes
spectrophotometry on 25~\AA\ centers from 5450--5650~\AA. \label{fig6}}
\end{figure}

While the absolute-flux zero-point of all \emph{HST} standards is tied to the
monochromatic value of flux for Vega of $3.46\times10^{-9}$~erg cm$^{-2}$
s$^{-1}$~\AA$^{-1}$ at 5556~\AA\ (Megessier 1995), there are valid absolute
measures in the IR, as summarized by Ri08. Because Ri08 discuss and present a
SED for Vega, the differences between our preferred Vega SED and the Ri08 SED
are investigated as a possible explanation for the differences between the
\emph{HST} and Ri08 flux calibrations for IRAC. The SED of Vega has been
measured by STIS at 0.17--1~$\mu$m by Bohlin \& Gilliland (2004), who suggested
that the Kurucz $T_{\rm eff}=9550$~K model\footnote{http://kurucz.harvard.edu/}
fits the STIS observations, while Bohlin (2007a) discovered that a change in the
STIS non-linearity correction made a Kurucz 9400~K model$^{4}$ fit the observed
flux much better both below the Balmer jump and in the 0.7--0.8~$\mu$m region,
as illustrated in Figure~\ref{fig6} with red for the 9550~K and green
for the 9400~K models. These two models and the reference continuum level are
all normalized to $3.46\times10^{-9}$~erg cm$^{-2}$ s$^{-1}$~\AA$^{-1}$ at
5556~\AA\ (Megessier 1995). The blue line is the same 9550~K model but is
normalized to the Ri08 IR value of 645Jy $\pm$2.3\% at 2.22~$\mu$m. The Ri08
fluxes below 1~\micron\ are not expected to match the actual stellar flux and
are not shown. The division of all the SEDs by the same smooth theoretical model
continuum in Figure~\ref{fig6} illustrates the differences among the
various flux distributions and also shows where the (mostly hydrogen) line
blanketing complicates the comparisons. The STIS measurements (black line in
Figure~\ref{fig6}) and the 9400~K model (green line) agree to $\sim$1\%
in the unblanketed regions from below the Balmer jump to $\sim$1~$\mu$m. 

Fortunately, the discrepancy of 1.1\% at 2.22~$\mu$m between the \emph{HST}
based fluxes and Ri08 is almost negligible and could be reduced by using the
9400~K model with a weighted average of the visible and IR normalizations.
Another complication is that Vega is a pole-on rapid rotator (Peterson et~al.
2006) with temperature zones in the 7900--10150~K range (Aufdenberg et~al.
2006). Even though the line blanketing in the Aufdenberg et~al.
multi-temperature model is occasionally stronger than measured, the continuum
levels of his computed pole-on SED (Aufdenberg, private comm.) track the ratio
of the 5556~\AA\  to the 2.22~\micron\ absolute fluxes to 0.4\%, i.e. much
better than any uncertainties in the absolute flux measurements in the visible
or IR. In summary, there is no reason to question the accuracy of either the
5556~\AA\ flux of Megessier (1995) or the 2.22~\micron\ flux quoted by Ri08.

A robust and straightforward comparison of a model SED for Vega with the
measured absolute visible and IR fluxes is not possible because of the IR
emission from the dust ring and because Vega is a pole-on rapid rotator. As
pointed out by Cohen et~al. (1992b), Sirius is a much better primary IR standard
because of its low rotation speed and lack of a contaminating dust ring. Thus,
one big step forward in resolving any possible offset between visible and IR
absolute fluxes is to observe Sirius with STIS and fit a model, in order to
directly compare the model with IR absolute fluxes, eg. from MSX (Price et~al.
2004), who have $\sim$1\% measurement accuracy in six bands from 4.3 to
21.3~\micron. STIS spectrophotometry of Sirius will be somewhat more saturated
than for Vega; but the same techniques of Bohlin \& Gilliland (2004) for
precisely analyzing saturated data should apply.

On longer time scales, there are on-going programs to measure stellar
spectrophotometric fluxes relative to NIST laboratory standards. The ACCESS
rocket program will establish a few fundamental flux standards in the brightness
range of Sirius to V$\sim$9.5 (Kaiser et~al. 2007). A ground-based program, NIST
STARS, is supported by lidar to measure the real-time atmospheric extinction and
uses NIST calibrated detectors to establish an all-sky set of SEDs for
standard stars with precisions of 0.5\% (Zimmer et~al. 2010).

\section{Summary}

The IRAC calibration constants have been determined from a somewhat different
set of flux standards than utilized by Re05. Small differences between our
results and those of Re05 are explained by differences in the photometry
extracted from the IRAC images. Our results are in agreement with the
independent IRAC4 8~\micron\ calibration of Ri08.

The robustness of our results suffer from a deficiency of G-star flux standards,
from the exclusion of two out of the four Re05 primary flux standards, and from
offsets of up to 4$\sigma$ between our set of WD stars and the cooler standards.
To alleviate these deficiencies, more comparison standards with \emph{HST} fluxes
and better model grids are required to model the measured fluxes and establish
SEDs to the JWST limit of 30~\micron.

\newpage 
\appendix{APPENDIX A} 

\section{\emph{HST} Calibration Methodology}

\subsection{Equations}

The \emph{HST} method of flux calibration for filter photometry does not involve
color corrections or nominal wavelengths and is always 
defined in terms of the photon weighted mean flux over the bandpass in
wavelength units
\begin{equation}\langle F_{\lambda}\rangle={\int F_\lambda~\lambda~R~d\lambda \over  
\int \lambda~R~d\lambda}=P_\lambda N_e(pred)\label{favl}\end{equation} 
or in frequency units
\begin{equation}\langle F_{\nu}\rangle={\int F_\nu~\nu^{-1}~R~d\nu \over \int 
\nu^{-1}~R~d\nu} \label{favnu}=P_\nu N_e(pred)\end{equation}
(Koornneef et~al. 1986, Ri08),  because detectors are generally photon detection
devices, rather than total energy sensing bolometers. Some authors, (e.g.
Bessel, et~al. 1998 \& Cohen, et~al. 2003), define our product $\lambda~R$  as
their response function of the system.

Source independent calibration constants $P$ are defined by dividing the mean
flux by the predicted electrons s$^{-1}$, $N_e(pred)$, in an infinite-radius
aperture. If the predicted $N_e(pred)$ from Equation (\ref{ne}) is substituted
in Equations (\ref{favl}-\ref{favnu}),   
\begin{equation}P_\lambda={hc \over
A\int{\lambda~R~d\lambda}} \label{pl}\end{equation} \begin{equation}P_\nu={h
\over A \int{\nu^{-1}~R~d\nu}}  \label{pnu}\end{equation}  
Because the \emph{HST} standard flux units are normally per unit wavelength, the
constant $P_\lambda$ appears in the headers of \emph{HST} photometric images
with the keyword name $photflam$. For some instruments, e.g. NICMOS, $P_\nu$
with the keyword name $photfnu$ is also included in the headers. The Re05
calibration constant in frequency units per Equations (\ref{fnu}) and (\ref{k})
is  \begin{equation}C'={F_{\nu_o}K\over N_e(pred)}=  {F_{\nu_o}  \over
N_e(pred)}{\int (F_\nu/F_{\nu_o}) (\nu/\nu_o)^{-1}~R~d\nu\over
{\int(\nu/\nu_o)^{-2}~R~d\nu}}~, \label{cp}\end{equation}  
After substituting
the definition of $\nu_o$ from Equation (\ref{lamo}) with
$\nu_o=c/\lambda_o$ and simplifying, the result is  $\langle
F_{\nu}\rangle/N_e(pred)$, i.e. the \emph{HST} and the Re05 methodologies
produce exactly the same calibration constants $C' \equiv P_\nu$.

The \emph{HST} calibration constants are normally derived from the source
independent Equations (\ref{pl}--\ref{pnu}) after any required adjustments are
made to the $R$ estimated from the product of laboratory component QE
measurements. These adjustments are derived by making the measured $N_e(obs)$ in
an infinite aperture match the predicted $N_e(pred)$ calculated from Equation
(\ref{ne}). In practice, a radius of something like the 5.5\arcsec\ for ACS is
defined as "infinite" (Sirianni et~al. 2005, Bohlin 2007b); and the primary pure
hydrogen WDs G191B2B, GD71, and GD153 are the preferred standards used for $F$
in Equation (\ref{ne}). Conversely for flux calibrated images as is the case for
the \emph{Spitzer} data, the measured $N_e(obs)$ can be calculated from the
measured mean-photometric flux as $N_e(obs)={\langle F_{\nu}\rangle / P_\nu}$.
This reconciliation of laboratory component throughputs versus the truth of
standard stars is achieved by adjusting the normalization of the filter
throughput or even by changing the quantum efficiency, QE, as function of
wavelength for the detector or filter when sufficient information exists (e.g.
de Marchi, et~al. 2004, Bohlin 2007b). Thus, information about individual
component throughputs, such as the telescope or detector QE, may be inferred
when there are multiple filters sampling the same wavelength regions.

To complement the above estimates of mean flux for stars imaged in a particular
filter, an associated wavelength is often useful. In addition to the nominal
wavelength $\lambda_o$ of Re05, other common definitions are the $mean$ and
$effective$ wavelengths.
\begin{equation}\lambda_{mean}={\int{\lambda~R~d\lambda} \over \int{
~R~d\lambda}}\label{lmean}\end{equation}
\begin{equation}\lambda_{eff}={\int{F_\lambda \lambda^2~R~d\lambda} \over
\int{F_\lambda \lambda ~R~d\lambda}} \label{leff}\end{equation} Perhaps, most
useful is the source independent \emph{pivot-wavelength} $\lambda_p$ and
associated \emph{pivot-frequency} $\nu_p$, where $\lambda_p \nu_p=c$ and 
$\langle F_{\lambda}\rangle~\lambda_p=\langle F_{\nu}\rangle~\nu_p$.
\begin{equation}\lambda_p=\sqrt{{c~\langle F_{\nu}\rangle}\over {\langle
F_\lambda\rangle}}= \sqrt{\int{\lambda~R~d\lambda}\over
\int{\lambda^{-1}~R~d\lambda}} \label{pivl}\end{equation} 
These various measures of the associated wavelength for a filter are in
Koornneef et~al. (1985) along with a definition of the rms width of a filter.
Appendix E of Ri08 contains our definition for mean wavelength but has an
alternative formula for a band width. The Ri08 definition of nominal
wavelength is photon weighted as in Re05 but is an integral over wavelength in
contrast to the Re05 integral over frequency. Also discussed in  Ri08 is the
concept of isophotal wavelength.

Having calculated the source independent \emph{pivot-wavelength} $\lambda_p$,
Equation (\ref{pivl}) provides a convenient formula for calculating $P_\lambda$
from $P_\nu=C'$ values.
\begin{equation}P_\lambda={c~P_\nu\over{\lambda_p}^2}\label{plam}\end{equation}

\subsection{Results}

One complication in the comparison of the \emph{HST} and Re05 flux calibration
methodologies is that the above discussion is for the flux calibration of a
point source, while the Spitzer images are calibrated in terms of surface
brightness I. If a calibration constants $C_I$ is defined so that multiplication
by the original DN/s produces I, then \begin{equation}P_\nu \equiv
C'={\Omega_{pix}C_I \over G}~,\label{pnucp}\end{equation}  where $\Omega_{pix}$
is the size of a pixel in steradians, G is the gain in electrons/DN, and the
data number (DN) is the unit of the instrumental signal.  Another small
complication is that the published values C of Re05 for the surface brightness
calibration are adjusted for an absolute calibration based on a ten pixel radius
for stellar photometry and require the correction to the infinite-aperture
$C_I=C*f_{10}$. Values for $f_{10}$ are listed in Re05 as 0.944, 0.937, 0.772,
and 0.737 for IRAC channels 1, 2, 3, and 4, respectively, and may have large
uncertainties because of complications caused by the small format and the
internal scattering in the Si:As array. These values appear in Table 5 along
with a selection of other calibration parameters, including \emph{photflam} and
\emph{photfnu}, the \emph{HST} style calibration constants. The top two rows of
Table 5 are the published calibration constants from the \emph{Spitzer}
instrumental references in our  \emph{Introduction}. The row labeled "R-Corr."
in Table 5 is the correction factor to the  published \emph{Spitzer} throughput
values R that brings the $N_e(pred)$ values calculated from Equation (\ref{ne})
into agreement with the measured $N_e(obs)=\langle F_{\nu}\rangle/P_\nu$~ for
the weighted average of the eight A and G stars used for IRAC. The $P_\nu$
values are from Re05 and applying the tabulated R-Corr to the published
throughput R curves makes the IRAC calibration self-consistent. Any change to an
adopted calibration constant requires a corresponding change to the absolute
level of that filter's R curve to maintain internal consistency. The R-Corr
factors have poor precision for IRSB with its linearity problem and for MIPS
with only three stars. Because the effective wavelength depends on the stellar
SED, Table 5 includes examples for our hottest star G191B2B and a cooler star
P330E; however, the differences are insignificant, because the slopes of the two
SEDs are nearly the same in  the IR. The flux weighting makes the
$\lambda_{eff}$ values smaller than the other measures of wavelength.

The IRSB relative response\footnote
{http://ssc.spitzer.caltech.edu/files/spitzer/bluePUtrans.txt}, R, has been
divided by 3.58 to make the IRSB system throughput comparable to the other five
modes and bring the values into agreement with Fig. 6.1 in the IRS Data
Handbook, where the peak QE is shown at a reasonable peak value of 0.81. Without
this adjustment of 3.58, the value of 0.54 for the IRSB throughput correction
in Table 5 would have been even smaller.

\acknowledgments

Support for this work was provided by NASA through the Space Telescope Science
Institute, which is operated by AURA, Inc., under NASA contract NAS5-26555.
Thanks to A. Gianninas for supplying the 60,000K SED with CNO that appears in 
Figure~\ref{fig3}.

\newpage

\begin{deluxetable*}{lrrrccc}
\tabletypesize{\scriptsize}

\tablewidth{0pt}

\tablecolumns{6}

\tablecaption{\emph{HST} and \emph{SPITZER} Comparison Stars}

\tablehead{
\colhead{Star} &\colhead{$R.A.$} &\colhead{$Decl.$} &\colhead{$V$} 
&\colhead{Sp.T} &\colhead{$T_\mathrm{eff}$}\\
&\multicolumn{1}{c}{J2000} &\multicolumn{0}{c}{J2000} &\multicolumn{0}{c}{(mag)}
&&\multicolumn{0}{c}{(K)}}

\startdata
GD71    &\phn~5 52 27.51 &+15 53 16.6 &13.032   &DA1  &32747\\
GD153     &12 57 02.37  &+22 01 56.0 &13.346   &DA1  &38686\\
G191B2B &\phn~5 05 30.62 &+52 49 54.0 &11.781   &DA0  &61193\\
LDS749B   &21 32 16.24  &+00 15 14.4 &14.674   &DBQ4  &13575\\
& \\
HD165459  &18 02 30.74  & 58 37 38.1 &\phn6.864 &A4V  &8600\\
1732526   &17 32 52.64  &+71 04 43.1 &12.530   &A4V  &8500\\
1740346   &17 40 34.7   & 65 27 15.0 &12.478   &A6V  &8050\\
1743045   &17 43 04.5   & 66 55 01.7 &13.525   &A8II  &7650\\
1802271   &18 02 27.17  &+60 43 35.6 &11.985   &A2V  &9100\\
1805292   &18 05 29.3   & 64 27 52.1 &12.278   &A4V  &8400\\
1812095   &18 12 09.57  & 63 29 42.3 &11.736   &A5V  &8250\\
& \\
HD209458  &22 03 10.78  &+18 53 03.7 &\phn7.65 &G0V  &6080\\
P041C     &14 51 58.19  &+71 43 17.3 &12.01    &G0V  &5960\\
P330E     &16 31 33.85  &+30 08 47.1 &13.01    &G0V  &5820\\
\enddata
\end{deluxetable*}

\begin{deluxetable*}{lccccc}
\tablewidth{0pt}
\tablecolumns{6}
\tablecaption{\emph{SPITZER} Calibration Parameters}
\tablehead{
\colhead{Channel} &\colhead{$\lambda_o$} &\colhead{Cal. Ap.} &\colhead{Ap. Corr.} 
&\colhead{Ap. Ref.\tablenotemark{a}} &\colhead{R Ref.\tablenotemark{b}}\\
&\multicolumn{1}{c}{(\micron)} &\multicolumn{0}{c}{(pixels)}}

\startdata
IRAC1  &3.544   &10      &1.112  &1  &3 \\
IRAC2  &4.487   &10      &1.113  &1  &3 \\
IRAC3  &5.710   &10      &1.125  &1  &3 \\
IRAC4  &7.841   &10      &1.218  &1  &3 \\
IRSB   &15.793  &$\infty$ &1.56   &2   &4 \\
MIPS24 &23.675  &$\infty$ &{\nodata\tablenotemark{c}}  &\nodata   &5 \\
\enddata
\tablenotetext{a}{References for aperture corrections: (1) Hora et~al.~(2008); 
(2)~IRS Instrument Handbook, 4.2.3.1 
(http://ssc.spitzer.caltech.edu/irs/irsinstrumenthandbook/45/)}
\tablenotetext{b}{References for the system-throughput spectral-response curves, R: 
(3) http://ssc.spitzer.caltech.edu/irac/calibrationfiles/spectralresponse/
(4) http://ssc.spitzer.caltech.edu/files/spitzer/bluePUtrans.txt divided by 3.58
(5) http://ssc.spitzer.caltech.edu/files/spitzer/MIPSfiltsumm.txt}
\tablenotetext{c}{PSF photometry is used for MIPS.}

\end{deluxetable*}

\clearpage

\begin{landscape}

\begin{deluxetable*}{lcccccccccccc}
\tabletypesize{\scriptsize}
\tablewidth{0pt}
\tablecaption{\emph{SPITZER} and Synthetic Photometry\tablenotemark{a} (mJy)}
\tablehead{
\colhead{Star} & 
 \multicolumn{2}{c}{IRAC1} &
 \multicolumn{2}{c}{IRAC2} &
 \multicolumn{2}{c}{IRAC3} &
 \multicolumn{2}{c}{IRAC4} &
 \multicolumn{2}{c}{IRSB} &
 \multicolumn{2}{c}{MIPS24} \\
& \colhead{Flux} & \colhead{Unc(\%)} &
 \colhead{Flux} & \colhead{Unc(\%)} &
 \colhead{Flux} & \colhead{Unc(\%)} &
 \colhead{Flux} & \colhead{Unc(\%)} &
 \colhead{Flux} & \colhead{Unc(\%)} &
 \colhead{Flux} & \colhead{Unc(\%)}}
\startdata
 G191B2B & 2.04e+00   & 0.64   & 1.29e+00   & 0.61   & 8.05e-01   & 1.24   & 4.08e-01   & 2.60   &  \nodata   &  \nodata   &  \nodata   &  \nodata  \\
         & 2.06e+00   &        & 1.31e+00   &        & 8.25e-01   &        & 4.57e-01   &        &  \nodata   &          &  \nodata    &   \\ 
   GD153 & 5.01e-01   & 0.70   & 3.12e-01   & 0.84   & 1.95e-01   & 4.16   & 1.07e-01   & 7.11   &  \nodata   &  \nodata   &  \nodata   &  \nodata  \\
         & 4.96e-01   &        & 3.13e-01   &        & 1.96e-01   &        & 1.08e-01   &        &  \nodata   &          &  \nodata    &   \\ 
    GD71 & 6.93e-01   & 0.69   & 4.39e-01   & 0.67   & 2.79e-01   & 3.26   & 1.45e-01   & 4.72   &  \nodata   &  \nodata   &  \nodata   &  \nodata  \\
         & 6.76e-01   &        & 4.25e-01   &        & 2.66e-01   &        & 1.45e-01   &        &  \nodata   &          &  \nodata    &   \\  
 LDS749B & 2.52e-01   & 0.84   & 1.59e-01   & 1.24   & 1.11e-01   & 6.08   & 5.56e-02   &12.34   &  \nodata   &  \nodata   &  \nodata   &  \nodata  \\
         & 2.65e-01   &        & 1.70e-01   &        & 1.07e-01   &        & 5.93e-02   &        &  \nodata   &          &  \nodata    &   \\ \\ 
HD165459 & 6.62e+02   & 0.08   & 4.26e+02   & 0.09   & 2.69e+02   & 0.17   & 1.48e+02   & 0.12   &  \nodata   &  \nodata   & 2.57e+01  & 0.29  \\
         & 6.50e+02   &        & 4.21e+02   &        & 2.69e+02   &        & 1.50e+02   &        &  \nodata   &          & 1.70e+01  &   \\ 
 1732526 & 3.65e+00   & 1.16   & 2.34e+00   & 1.74   & 1.50e+00   & 3.70   &  \nodata    &  \nodata & 1.58e-01 & 6.03      &  \nodata   &  \nodata  \\
         & 3.58e+00   &        & 2.32e+00   &        & 1.48e+00   &        &  \nodata    &        & 2.10e-01  &          &  \nodata    &   \\ 
 1740346 & 4.62e+00   & 0.86   & 3.02e+00   & 1.13   & 1.97e+00   & 1.78   & 1.12e+00   & 1.00   & 6.59e-01 & 2.93      &  \nodata   &  \nodata  \\
         & 4.45e+00   &        & 2.89e+00   &        & 1.85e+00   &        & 1.03e+00   &        & 2.64e-01 &          &  \nodata    &  \\ 
 1743045 & 2.13e+00   & 0.22   & 1.38e+00   & 0.27   & 8.88e-01   & 0.43   & 4.83e-01   & 0.41   &  \nodata   &  \nodata   &  \nodata   &  \nodata  \\
         & 2.07e+00   &        & 1.34e+00   &        & 8.59e-01   &        & 4.77e-01   &        &  \nodata   &          &  \nodata   &   \\ 
 1802271 & 5.19e+00   & 0.26   & 3.35e+00   & 0.31   & 2.16e+00   & 0.58   & 1.16e+00   & 0.36   & 2.26e-01  & 6.55    &  \nodata    &  \nodata  \\
         & 5.12e+00   &        & 3.32e+00   &        & 2.12e+00   &        & 1.17e+00   &        & 3.00e-01  &         &  \nodata    &  \\ 
 1805292 & 4.58e+00   & 0.95   & 2.99e+00   & 1.09   & 1.93e+00   & 3.55   & 1.07e+00   & 1.07   & 2.28e-01  & 4.61    &  \nodata    &  \nodata  \\
         & 4.43e+00   &        & 2.88e+00   &        & 1.84e+00   &        & 1.02e+00   &        & 2.62e-01  &         &  \nodata    &  \\  
 1812095 & 8.85e+00   & 0.07   & 5.75e+00   & 0.09   & 3.66e+00   & 0.19   & 1.98e+00   & 0.16   & 4.96e-01  & 4.68    & 2.04e-01  & 8.82  \\
         & 8.61e+00   &        & 5.59e+00   &        & 3.58e+00   &        & 1.99e+00   &        & 5.10e-01  &         & 2.26e-01  &   \\
HD209458\tablenotemark{b} & \nodata & \nodata & \nodata & \nodata & \nodata  & \nodata  & \nodata     & \nodata  & \nodata    & \nodata   & 2.13e+01  & 0.11  \\
                      & \nodata &       &  \nodata &       & \nodata  &         & \nodata     &        & \nodata    &          & 2.21e+01  &       \\ 
   P041C & 1.79e+01   & 0.77   & 1.12e+01   & 0.97   & 7.39e+00   & 1.56   & 4.12e+00   & 2.07   & 1.14e+00  & 6.64    &  \nodata   &  \nodata  \\
         & 1.73e+01   &        & 1.10e+01   &        & 7.09e+00   &        & 3.98e+00   &        & 1.02e+00  &         &  \nodata   &  \\
   P330E & 7.84e+00   & 0.73   & 5.07e+00   & 0.88   & 3.21e+00   & 3.43   & 1.80e+00   & 1.65   &  \nodata   &  \nodata  & 2.20e-01  & 5.45  \\
         & 7.78e+00   &        & 4.97e+00   &        & 3.20e+00   &        & 1.79e+00   &        &  \nodata   &          & 2.04e-01  &   \\
\enddata
\tablenotetext{a}{The first line for each star contains the measured fluxes,
while the second line is the predicted synthetic photometry.}
\tablenotetext{b}{Existing IRAC and IRSB data are only in non-standard modes.}
\end{deluxetable*}

\clearpage
\end{landscape}

\begin{deluxetable}{lccc}
\tabletypesize{\scriptsize}
\tablewidth{0pt}
\tablecolumns{4}
\tablecaption{${C'_{ST}/C'}$=~New/Re05 IRAC Calibration Constants}
\tablehead{
\colhead{Channel} &\colhead{Star Set} &\colhead{New/Orig.} &\colhead{Unc.}}
\startdata
IRAC1  & 4 WD     &1.004  &0.006\\
IRAC1  & 2 Primary &0.977  &0.007\\
IRAC1  & 8 A+G    &0.977  &0.004\\
& \\
IRAC2  & 4 WD	  &1.006  &0.006\\
IRAC2  & 2 Primary &0.981  &0.007\\
IRAC2  & 8 A+G  &0.981  &0.004\\
& \\
IRAC3  & 4 WD	  &1.007  &0.013\\
IRAC3  & 2 Primary &0.988  &0.007\\
IRAC3  & 8 A+G    &0.980  &0.005\\
& \\
IRAC4  & 4 WD	  &1.077  &0.024\\
IRAC4  & 2 Primary &1.005  &0.007\\
IRAC4  &  7 A+G   &0.995  &0.005\\
\enddata

\end{deluxetable}

\begin{landscape}

\begin{deluxetable}{cccccccc}
\tabletypesize{\scriptsize}
\tablewidth{0pt}
\tablecolumns{8}
\tablecaption{\emph{HST} Calibration Methodology for Absolute Flux on the Re05
Scale}
\tablehead{
\colhead{Item} &\colhead{Units} &\colhead{IRAC1} &\colhead{IRAC2}
&\colhead{IRAC3} &\colhead{IRAC4} &\colhead{IRSB} &\colhead{MIPS24}}
\startdata
C (Re05 10px)&	  $MJy~sr^{-1}/(DN~s^{-1})$ &0.1088 &0.1388 &0.5952 &0.2021 &\nodata &\nodata \\
$C_I$		  &$MJy~sr^{-1}/(DN~s^{-1})$ &0.1027 &0.1301 &0.4595 &0.1489 &0.0117\tablenotemark{a} &0.0454\\
$f_{10}$           &    &0.944 &0.937 &0.772 &0.737  &\nodata &\nodata \\
Gain	  	  &electrons/DN  &3.30  &3.71  &3.80  &3.80  &\nodata  &5 \\
$\Omega_{pix}$	  &sr  &3.498~$10^{-11}$ &3.498~$10^{-11}$  &3.498~$10^{-11}$  &3.498~$10^{-11}$ &7.893~$10^{-11}$ &1.457~$10^{-10}$\\
$P_\nu$=C' 	  &$MJy/(e~s^{-1})$ &1.089e-12 &1.226e-12 &4.230e-12 &1.371e-12 &9.235e-13 &1.323e-12 \\
$P_\lambda$	  &$erg~s^{-1}~cm^{-2}~\AA^{-1}/(e~s^{-1})$ &2.589e-20 &1.819e-20 &3.870e-20 &6.613e-21  &1.093e-21 &7.026e-22 \\
R-Corr	  	  &  &1.400 &1.071 &1.050 &0.985 &0.540 &0.485 \\
$\lambda_o$	  &$\micron$  &3.544 &4.487 &5.710 &7.841 &15.793 &23.675 \\
$\lambda_p$	  &$\micron$  &3.551 &4.496 &5.724 &7.884 &15.916 &23.759 \\
$\lambda_{mean}$  &$\micron$  &3.557 &4.505 &5.739 &7.927 &16.040 &23.843 \\
$\lambda_{eff}$(G191B2B)  &$\micron$  &3.519 &4.453 &5.656 &7.675 &15.417 &23.374 \\
$\lambda_{eff}$(P330E)	  &$\micron$  &3.521 &4.452 &5.659 &7.674 &15.403 &23.361 \\
\enddata
\tablenotetext{a}{Units are per electron rather than DN, so a gain factor is not needed.}

\end{deluxetable}

\clearpage
\end{landscape}

\end{document}